\documentclass[preprint,notoc]{JHEP3}
\usepackage{epsfig}

\def\to{\rightarrow}

\def\bi{\begin{itemize}}
\def\ei{\end{itemize}}
\def\te{\tilde e}

\def\tu{\tilde u}

\def\tb{\tilde b}

\def\td{\tilde d}

\def\tst{\tilde t}
\def\ttau{\tilde \tau}

\def\tg{\tilde g}
\def\tnu{\tilde\nu}
\def\tell{\tilde\ell}

\def\tw{\widetilde W}
\def\tz{\widetilde Z}

\def\agt{\stackrel{>}{\sim}}

\title{Updated Constraints on the \\
Minimal Supergravity Model}

\author{Howard Baer, Csaba Bal\'azs, Alexander Belyaev\footnote
{On leave of absence from Nuclear Physics Institute, Moscow State University.} 
\\ Department of Physics, Florida State University\\ 
Tallahassee, FL, USA 32306\\
E-mail: \email{baer@hep.fsu.edu}, \email{balazs@hep.fsu.edu},
        \email{belyaev@hep.fsu.edu}}
\author{J.~Kenichi Mizukoshi
\\
Instituto de F\'{\i}sica, Universidade de S\~ao Paulo \\
C.P.\ 66318, 05315--970, S\~ao Paulo, Brazil. \\
E-mail: \email{mizuka@fma.if.usp.br}} 
\author{Xerxes Tata, Yili Wang \\
Department of Physics and Astronomy, University of Hawaii, \\
Honolulu, HI 96822, USA    \\
E-mail: \email{tata@phys.hawaii.edu}, \email{yili@phys.hawaii.edu}}

\preprint{\vbox{\hbox{FSU-HEP-020530} \vspace{0.2cm}
                \hbox{IFUSP-1549/2002} \vspace{0.2cm}
                \hbox{UH-511-1002-02}}} 

\abstract{Recently, refinements have been made on both the
theoretical and experimental determinations of 
{\it i}.) the mass of the lightest Higgs scalar ($m_h$),
{\it ii}.) the relic density of cold dark matter in the 
universe ($\Omega_{CDM}h^2$),
{\it iii}.) the branching fraction for radiative
$B$ decay $BF( b\to s \gamma )$,  
{\it iv}.)  the muon anomalous magnetic moment ($a_\mu$), and
{\it v}.) the flavor violating decay $B_s\to \mu^+\mu^-$. 
Each of these quantities can be predicted in the MSSM, and each depends 
in a non-trivial way on the spectra of SUSY particles.
In this paper, we present updated constraints from each of these quantities
on the minimal supergravity (mSUGRA) model as embedded in the
computer program ISAJET. The combination of constraints
points to certain favored regions of model parameter space where
collider and non-accelerator SUSY searches may be more focussed.
}


\begin{document}

\section{Introduction}
\label{sec:intro}

The search for weak scale supersymmetric matter is one of the prime objectives
of present and future collider experiments. Particle physics models including
supersymmetry solve a host of problems occurring in non-supersymmetric
theories, and predict a variety of new matter states--- the sparticles---
at or around the TeV scale\cite{reviews}. 
Supersymmetric models can be classified
by the mechanism for communicating SUSY breaking from the hidden
sector to the observable sector. Possibilities include gravity
mediated SUSY breaking (SUGRA)\cite{sugra}, 
gauge mediated SUSY breaking (GMSB)\cite{gmsb},
anomaly mediated SUSY breaking (AMSB)\cite{amsb} and gaugino mediated SUSY
breaking (inoMSB)\cite{inomsb}. Of these, the SUGRA models may be perceived as
the most conservative, since they do not require the introduction
of either extra dimensions or new messenger fields, and because gravity
exists. 

The so-called {\it minimal} supergravity (mSUGRA) model (sometimes also
referred to as the CMSSM model) has traditionally been the most popular
choice for phenomenological SUSY analyses. In mSUGRA, it is assumed that
the Minimal Supersymmetric Standard Model (MSSM) is valid from the weak
scale all the way up to the GUT scale $M_{GUT}\simeq 2\times 10^{16}$
GeV, where the gauge couplings $g_1$ and $g_2$ unify. In many of the
early SUGRA models\cite{sugra}, a simple choice
of K\"ahler metric $G_i^j$ and gauge kinetic function $f_{AB}$ led to
{\it universal} soft SUSY breaking scalar masses ($m_0$), gaugino masses
($m_{1/2}$) and $A$-terms ($A_0$) at $M_{GUT}$.  This assumption of
universality in the scalar sector
leads to the phenomenologically required suppression of
flavor violating processes that are supersymmetric in origin.  However,
there is no known physical principle which gives rise to the desired
form of $G_i^j$ and $f_{AB}$; indeed, for general forms of $G_i^j$ and
$f_{AB}$, non-universal masses are expected\cite{sw}.  In addition, even
if nature did select a SUGRA model leading to tree level universality,
quantum corrections would (without further assumptions) lead to large
deviations from universality\cite{munoz}.  Hence, the universality
assumption nowadays is regarded as being ad hoc--- entirely motivated by
the phenomenological need for suppression of flavor violating processes
in the MSSM.

In the mSUGRA model, we thus assume universal scalar masses, gaugino
masses and $A$-terms. We will also require that electroweak symmetry is
broken radiatively (REWSB), allowing us to fix the magnitude,
but not the sign, of the superpotential Higgs mass term $\mu$ so as to
obtain the correct value of $M_Z$. Finally, we
trade the bilinear soft
supersymmetry breaking (SSB) parameter $B$ for $\tan\beta$ (the ratio of
Higgs field vacuum expectation values).  Thus,
the parameter set
\begin{equation}
m_0,\ m_{1/2},\ A_0,\ \tan\beta ,\ \ {\rm and}\ \ sign(\mu )
\end{equation}
completely determines the spectrum of supersymmetric matter and Higgs fields.

In our calculations, we use the program ISASUGRA to calculate the SUSY
particle mass spectrum. ISASUGRA is part of the ISAJET\cite{isajet}
package. Working in the $\overline{DR}$ regularization scheme, the
weak scale values of gauge and third generation Yukawa couplings
are evolved via 2-loop RGEs to $M_{GUT}$. At $M_{GUT}$, universal
SSB boundary conditions are imposed, and all SSB masses along with gauge 
and Yukawa couplings are evolved to the weak scale $M_{weak}$. 
Using an optimized scale choice $Q_{SUSY}=\sqrt{m_{\tst_L}m_{\tst_R}}$, 
the RG-improved one-loop effective potential is minimized
and the entire spectrum of SUSY and Higgs particles is calculated.
Our values of $m_h$ are in close accord with those generated by the
FeynHiggsFast program\cite{FHF}.
Yukawa couplings are updated\cite{bagger} 
to account for SUSY threshold corrections,
and the entire parameter set is iteratively run between $M_{weak}$ and
$M_{GUT}$ until a stable solution (within tolerances) is obtained.

Once the SUSY and Higgs masses and mixings are known, then a host of
observables may be calculated, and compared against experimental
measurements. The most important of these include:
\begin{itemize}
\item lower limits on sparticle and Higgs boson masses from new particle
searches at LEP2,
\item the relic density of neutralinos originating from the Big Bang,
\item the branching fraction of the flavor changing decay $b\to
s\gamma$,
\item the value of muon anomalous magnetic moment $a_\mu = \frac{(g-2)_\mu}{2}$ and
\item the lower bound on the rate for the rare decay $B_s\to\mu^+\mu^-$.
\end{itemize}

Our goal is to delineate the mSUGRA parameter space region consistent
with all these constraints.  Similar studies have recently been
presented in Refs. \cite{Gomez:2002tj,ellis,leszek,drees,deboer}.  
In this report we
examine these constraints using the updated ISASUGRA package, which is
convenient for explicit event generation for various colliders using the
ISAJET v7.63 program, which includes several improvements over previous
versions. The most important of these improvements are the evaluation of
the bottom Yukawa coupling and the value of $m_A$, especially for large
values of $\tan\beta$. Complete 1-loop self energy corrections as given
in Ref. \cite{bagger} have been incorporated.\footnote{On the technical
side, the numerical precision in ISAJET has been improved to facilitate
a better analysis near the boundary of the region excluded by
electroweak symmetry breaking constraints.}  
Comparisons of ISAJET v7.63 with other similar codes are available 
in Ref. \cite{kraml}.

Returning to our analysis,
we also incorporate a new calculation of the neutralino relic density
$\Omega_{\tz_1}h^2$ that has recently become available\cite{bbb}.  In
Ref. \cite{bbb}, all relevant neutralino annihilation and
co-annihilation processes are calculated, and the neutralino relic
density is evaluated using {\it relativistic} thermal averaging (see
also Refs. \cite{bb} and recently Belanger {\it et al.},
Ref. \cite{belanger}). The latter is especially important in evaluating
the relic density when $s$-channel annihilation resonances occur, as in
the mSUGRA model at large $\tan\beta$, when $\tz_1\tz_1\to A,\ H\to
f\bar{f}$, where the $f$s are SM fermions\cite{bb,dn,an,ellis_ltb,nano}. We
also present improved $b\to s\gamma$ branching fraction predictions in
accord with the current ISAJET release. We also discuss constraints
imposed by the measurement of the muon anomalous magnetic moment, using updated
calculations from two somewhat different analyses. Finally, we delineate
the region of mSUGRA parameter space excluded by the CDF lower
limit\cite{cdf} on the branching fraction of $B_s\to \mu^+\mu^-$ .  This
constraint is important only for very large values of
$\tan\beta$\cite{bmm}.

Within the mSUGRA framework, the parameters $m_0$ and $m_{1/2}$ are the
most important for fixing the scale of sparticle masses. The
$m_0-m_{1/2}$ plane (for fixed values of other parameters) is convenient
for
a simultaneous display of these constraints, and hence, of parameter
regions in accord with all experimental data.
Physicists interested in the mSUGRA model may 
wish to focus their attention on these regions.
We also present five mSUGRA model cases illustrating distinctive 
characteristics of the mSUGRA particle spectrum for parameter choices which 
are consistent with all experimental constraints.
  
The remainder of this paper is organized as follows. 
In Sec. \ref{sec:constraints}, we discuss the various constraints
on the mSUGRA model, and present some details of our calculations. 
In Sec. \ref{sec:results}, we show our main results as regions of the
$m_0\ vs.\ m_{1/2}$ parameter space plane for different values
of $\tan\beta$ and sign of $\mu$. Our conclusions and sample
points are presented in Sec. \ref{sec:conclude}.

\section{Constraints and calculations in the mSUGRA model}
\label{sec:constraints}

\subsection{Constraints from LEP2 searches}

The LEP2 collaborations have finished taking data, and
significant numbers of events were recorded at 
$e^+e^-$ CM energies ranging up to $\sqrt{s}\simeq 208$ GeV.
Based on negative searches for superpartners at LEP2,
we require 
\begin{itemize}
\item $m_{\tw_1}>103.5$ GeV\cite{lep2_w1} and
\item $m_{\te_{L,R}}>99$ GeV provided $m_{\tell}-m_{\tz_1}<10$ 
GeV\cite{lep2_sel}, which is the most stringent of the slepton mass
limits.
\end{itemize}

The LEP2 experiments also searched for the SM Higgs boson. In addition to 
finding several compelling signal candidates consistent with 
$m_h\sim 115$ GeV, they set a limit
$m_{H_{SM}}>114.1$ GeV\cite{lep2_h}. In our mSUGRA parameter space scans, the
lightest SUSY Higgs boson $h$ is almost always SM-like. The exception occurs
when the value of $m_A$ becomes low, less than $100-150$ GeV. Then there
exists a near mass degeneracy between $h$ and $H$, and the SM
Higgs is a mixture of these. This case arises at very large values 
of $\tan\beta$. For clarity, we show contours where
\bi
\item $m_h>114.1$ GeV,
\ei
and will direct the reader's attention to any regions where this bound might
fail.

\subsection{Neutralino relic density}

Measurements of galactic rotation curves, binding of galactic clusters, 
and the large scale structure of the universe all point to the need for
significant amounts of cold dark matter (CDM) in the universe. In addition, 
recent measurements of the power structure of the cosmic
microwave background, and measurements of distant supernovae, point
to a cold dark matter density\cite{cdm}
\bi
\item $0.1 <\Omega_{CDM}h^2<0.3$ .
\ei
The lightest neutralino of mSUGRA is an excellent candidate for
relic CDM particles in the universe.
The upper limit above represents a true constraint, while the 
corresponding lower
limit is flexible, since there may be additional sources of CDM such
as axions, or states associated with the hidden sector and/or extra
dimensions.

To estimate the relic density of neutralinos in the mSUGRA model,
we use the recent calculation in Ref. \cite{bbb}. In Ref. \cite{bbb},
all relevant neutralino annihilation and co-annihilation
reactions are evaluated at tree level using the CompHEP\cite{comphep}
program for automatic evaluation of the 
associated $7618$ Feynman diagrams. The 
annihilation cross section times velocity is relativistically thermally
averaged\cite{graciela}, 
which is important for obtaining the correct neutralino relic 
density in the vicinity of annihilations through $s$-channel resonances.

\subsection{The $b\to s\gamma$ branching fraction}

The branching fraction $BF(b\to s\gamma )$ has recently been measured by
the BELLE\cite{belle}, CLEO\cite{cleo} and ALEPH\cite{aleph}
collaborations.  Combining statistical and systematic errors in
quadrature, these measurements give $(3.36\pm 0.67)\times 10^{-4}$
(BELLE), $(3.21\pm 0.51)\times 10^{-4}$ (CLEO) and $(3.11\pm 1.07)\times
10^{-4}$ (ALEPH). A weighted averaging of these results yields $BF(b\to
s\gamma )=(3.25\pm 0.37) \times 10^{-4}$. The 95\% CL range corresponds
to $\pm 2\sigma$ away from the mean. To this we should add uncertainty
in the theoretical evaluation, which within the SM dominantly comes from
the scale uncertainty, and is about 10\%.\footnote{We caution the reader
that the SUSY contribution may have a larger theoretical uncertainty,
particularly if $\tan\beta$ is large. An additional theoretical
uncertainty that may increase the branching ratio in the SM is pointed
out in Ref. \cite{Gambino:2001ew}.} Together, these imply the bounds,
\bi
\item $2.16\times 10^{-4}< BF(b\to s\gamma )< 4.34 \times 10^{-4}$ .
\ei
Other computations of the range of $BF(b\to s\gamma )$ include for instance
Ellis {\it et al.}\cite{ellis}: 
$2.33\times 10^{-4}<BF(b\to s\gamma )<4.15\times 10^{-4}$,
and Djouadi {\it et al.}\cite{drees}: 
$2.0\times 10^{-4}<BF(b\to s\gamma )<5.0 \times 10^{-4}$.
In our study, we simply show contours of $BF(b\to s\gamma )$ of
2, 3, 4 and $5\times 10^{-4}$, allowing the reader the flexibility
of their own interpretation.

The calculation of $BF(b\to s\gamma )$ used here is based upon the
program of Ref. \cite{bsg}. That calculation uses an effective field 
theory approach to evaluating radiative corrections to the 
$b\to s\gamma$ decay rate. In running from $M_{GUT}$ to $M_{weak}$,
when any sparticle threshold is crossed, the corresponding sparticle is 
integrated out of the theory, and a new basis of decay-mediating operators
multiplied by Wilson coefficients (WCs) is induced. 
The evolution of the WCs can be calculated by RG methods. We adopt the
Anlauf procedure\cite{anlauf} in our calculation, which implements a tower 
of effective field theories, corresponding to each sparticle threshold
which is crossed. 
This procedure sums large logarithms that can occur 
from a disparity between different scales involved in the loop 
calculations. 
In our calculations, we implement the running $b$-quark mass
including SUSY threshold corrections as calculated in ISASUGRA;
these effects can be important at large values of the 
parameter $\tan\beta$\cite{degrassi,carena}.
Once the relevant operators and Wilson coefficients are 
known at $Q=M_W$, then the SM WCs are evolved down to $Q=m_b$ via
NLO RG running. At $m_b$, the $BF(b\to s\gamma )$ is evaluated at NLO,
including bremsstrahlung effects. Our value of the SM $b\to s\gamma$
branching fraction yields $3.4\times 10^{-4}$, with a scale uncertainty
of 10\%.     

\subsection{Muon anomalous magnetic moment}

The muon anomalous magnetic moment $a_\mu =\frac{(g-2)_\mu}{2}$ has been
recently measured to high precision by the E821 experiment\cite{e821}:
$ a_\mu=11659202(14)(6)\times 10^{-10}$.
In addition, additional data analyses should soon be finished, and we
may anticipate a further reduction in the experimental error by a factor
of 2.  The initially reported\cite{e821} $2.6\sigma$ deviation from the
SM value of Ref.\cite{marciano} has since been tempered somewhat by
correcting the sign of the SM light-by-light contribution to
$a_\mu$\cite{lbl}. A correction of the sign of the hadronic light by
light contribution reduces the significance of the deviation from the SM
to $1.6\sigma$, {\it i.e.}  at $2\sigma$:  
\bi
\item $-6<\delta a_\mu\times 10^{10}<58$ (CM). 
\ei
An alternative evaluation of theory uncertainties in the SM $a_\mu$
calculation by Melnikov\cite{melnikov} leads to (including the LBL
correction):
\bi
\item $-29.9< \delta a_\mu\times 10^{10}<62.3$ (Melnikov).
\footnote{Melnikov--- who uses the analysis of Ref.\cite{jag}
which does not use tau decay data for the evaluation of the hadronic
vacuum polarization and has a more conservative error on the light by
light contribution--- finds $\delta a_{\mu}\times 10^{10}= 16.2 \pm
14.0|_{stat} \pm 6.0|_{sys} \pm 15.6|_{theory}$. The conservative
``$2\sigma$'' range reported here can be obtained by linearly combining
the theory error with the $2\sigma$ experimental error.} 
\ei
In view of the
theoretical uncertainty, and the impending new experimental analysis, we only
present contours of $\delta a_\mu$, as calculated using the program
developed in \cite{bbft}, and leave it to the reader to decide the
extent of the parameter region allowed by the data. 

\subsection{$B_s\to\mu^+\mu^-$ decay}

While all SUSY models contain two doublets of Higgs superfields, there
are no tree level flavor changing neutral currents because one doublet
${\hat H}_u$ couples only to $T_3=1/2$ fermions, while the other doublet
$\hat{H}_d$ couples just to $T_3= -1/2$ fermions. At one loop, however,
couplings of ${\hat H}_u$ to down type fermions are induced. These
induced couplings grow with $\tan\beta$. As a result, down quark Yukawa
interactions and down type quark mass matrices are no longer
diagonalized by the same transformation, and flavor violating couplings
of neutral Higgs scalars $h$, $H$ and $A$ emerge. Of course, in the limit of
large $m_A$, the Higgs sector becomes equivalent to the SM
Higgs sector with the light Higgs boson $h=H_{SM}$, and the flavor
violation decouples. The interesting thing is that while
this decoupling occurs as $m_A \to \infty$, {\it there is no decoupling
for sparticle masses becoming large.}

An important consequence of this coupling is the possibility of the
decay $B_s \to \mu^+\mu^-$, whose branching fraction has been
experimentally
bounded by CDF\cite{cdf} to be:
\bi
\item $BF(B_s\to\mu^+\mu^- )< 2.6\times 10^{-6}$, 
\ei
mediated by the neutral states in the Higgs sector of supersymmetric
models. While this branching fraction is very small within the SM
($BF_{SM}(B_s \to \mu^+\mu^-)\simeq 3.4 \times 10^{-9}$), the amplitude
for the Higgs-mediated decay of $B_s$ grows as $\tan^3\beta$ within the
SUSY framework, and hence can completely dominate the SM contribution if
$\tan\beta$ is large.  Several groups\cite{bmm} have analyzed the
implications of this decay within the mSUGRA framework. A subset of
us\cite{mtw} have recently performed an independent analysis of this
decay. In the following, we use the results of this analysis to
delineate the region of mSUGRA parameters excluded by the CDF upper
limit on its branching fraction. 

Tevatron experiments should be able to probe this decay in the near
future.  With an integrated sample of 2~fb$^{-1}$ they should be
sensitive to a branching fraction for $B_s \to \mu^+\mu^-$ down to $\sim
10^{-7}$. With a still bigger data sample (that is expected to accumulate
before the Large Hadron Collider begins operation) the sensitivity
should be even greater.

\section{Results}
\label{sec:results}

Our first results are plotted in Fig. \ref{fig:sug10m}. Here, we
show the $m_0\ vs.\ m_{1/2}$ plane for $A_0=0$, $\tan\beta =10$
and $\mu <0$.
The red shaded regions are excluded either due to a
lack of REWSB (right-hand side), or a stau LSP (left-hand side).
The magenta region is excluded by searches  for charginos and
sleptons at LEP2. The region below the red contour is excluded by LEP2
Higgs searches, since here $m_h<114.1$ GeV. In addition, we show
regions of neutralino relic density with 
$0.1 <\Omega_{\tz_1} h^2 <0.3$ (green),
which is favored by cosmological observations, and also
$0.3 <\Omega_{\tz_1} h^2 <1$ (blue) and 
$0.02 <\Omega_{\tz_1} h^2 <0.1$ (yellow). The 
bulk of the unshaded region in the center of the plot has 
$\Omega_{\tz_1} h^2>1$, 
and would thus be excluded since the age of the universe would be less
than 10 billion years. The magenta contours denote values of
$BF(b\to s\gamma )= 4$ and $5\times 10^{-4}$. Finally, the blue
contours denote values of $\delta a_\mu =-30, -10, -5$ and 
$-2\times 10^{-10}$, moving from lower left to upper right.
There is no constraint arising from $B_s\to \mu^+\mu^-$ decay at 
$\tan\beta =10$.

An intriguing feature of the plot is that the lower left green shaded
region of desirable relic density, where neutralinos mainly annihilate
via $t$-channel slepton exchange 
to lepton-anti-lepton pairs
is essentially excluded by the $m_h$, $b\to s\gamma$ and
$\delta a_\mu$ constraints. That leaves two allowed regions
with a preferred relic density: one runs near the stau LSP region, where
$\ttau_1-\tz_1$ co-annihilation effects reduce an otherwise
large relic density 
(as pointed out by Ellis {\it et al.}\cite{ellis_co}).
This region has a highly fine-tuned relic density, since a slight change in
$m_0$ leads to either too light or too heavy of a $\ttau_1$ mass
to give $0.1<\Omega h^2<0.3$\cite{ellis_ft,bbb}. 
The other region runs parallel
to the REWSB excluded region, and occurs when the $\tz_1$ has a sufficiently
large higgsino component that annihilation into $WW$, $ZZ$ and $Zh$
pairs reduces the relic density\cite{feng_relic,bbb}. 
This region corresponds to what is known
as ``focus point'' SUSY, and since $m_0$ is large, SUSY scalar masses
are also large, leading to some degree of suppression of FCNC and CP violating 
processes\cite{feng}. The narrowness of the region indicates again
that some fine-tuning of parameters is needed to achieve the 
right relic density, although the amount of fine-tuning is less than
in the $\ttau_1$ co-annihilation case.

\FIGURE[t]{\epsfig{file=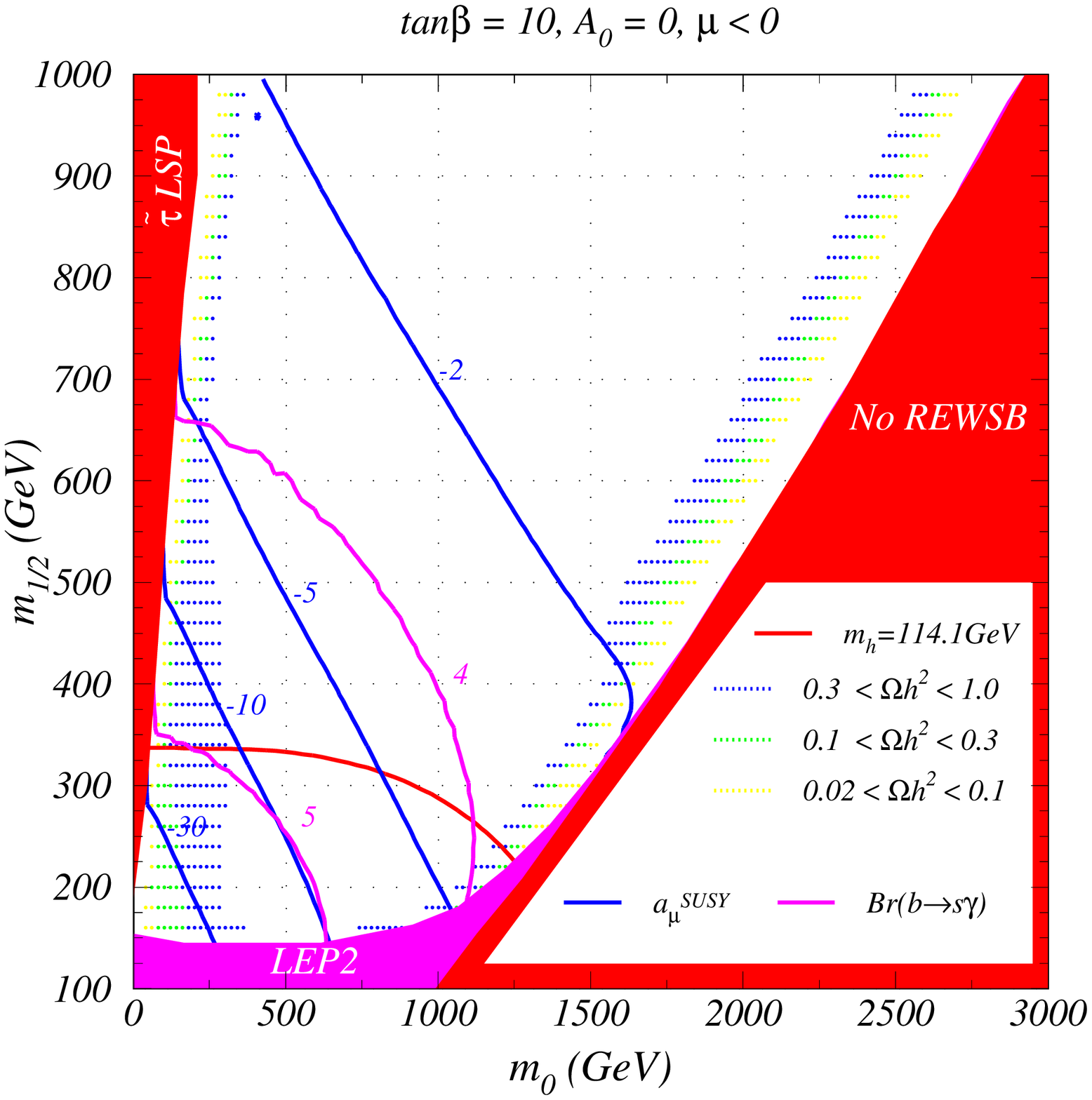,width=15cm} 
\caption{Plot of constraints for the mSUGRA model in the 
$m_0\ vs.\ m_{1/2}$ plane for $\tan\beta =10$, $A_0=0$ and $\mu <0$.
We show regions of CDM relic density, plus contours of $m_h=114.1$ GeV, 
contours of muon anomalous magnetic moment $a_\mu$ ($\times 10^{10}$) and
contours of $b\to s\gamma$ branching fraction ($\times 10^{4}$).}
\label{fig:sug10m}}

A similar plot is shown in Fig. \ref{fig:sug10p}, but in this case for
$\mu >0$. Much of the labeling is similar to Fig. \ref{fig:sug10m},
although now the $b\to s\gamma$ contour denotes a branching fraction of
$3\times 10^{-4}$. In this case almost the entire plane shown is in
accord with the measured branching fraction for this decay. In addition,
the blue contours denote values of $\delta a_\mu =60$, 40, 20, 10, 5 and
$2\times 10^{-10}$. Constraints from $\delta a_{\mu}$ as
well as from $B_s \to \mu^+\mu^-$ are not relevant for this case.

For $\tan\beta =10$ and $\mu >0$, the slepton annihilation region
of relic density has a small surviving region just beyond 
the Higgs mass contour. 
For the most part, to attain a 
preferred value of neutralino relic density, one must again live in the 
stau co-annihilation region, or the focus point region. A final 
possibility is to be in the slepton annihilation region, but then the 
value of $m_h$ should be slightly beyond the LEP2 limit; 
in this case,
a Higgs boson signal may be detected in Run 2 of the Fermilab
Tevatron\cite{run2}.

\FIGURE[t]{\epsfig{file=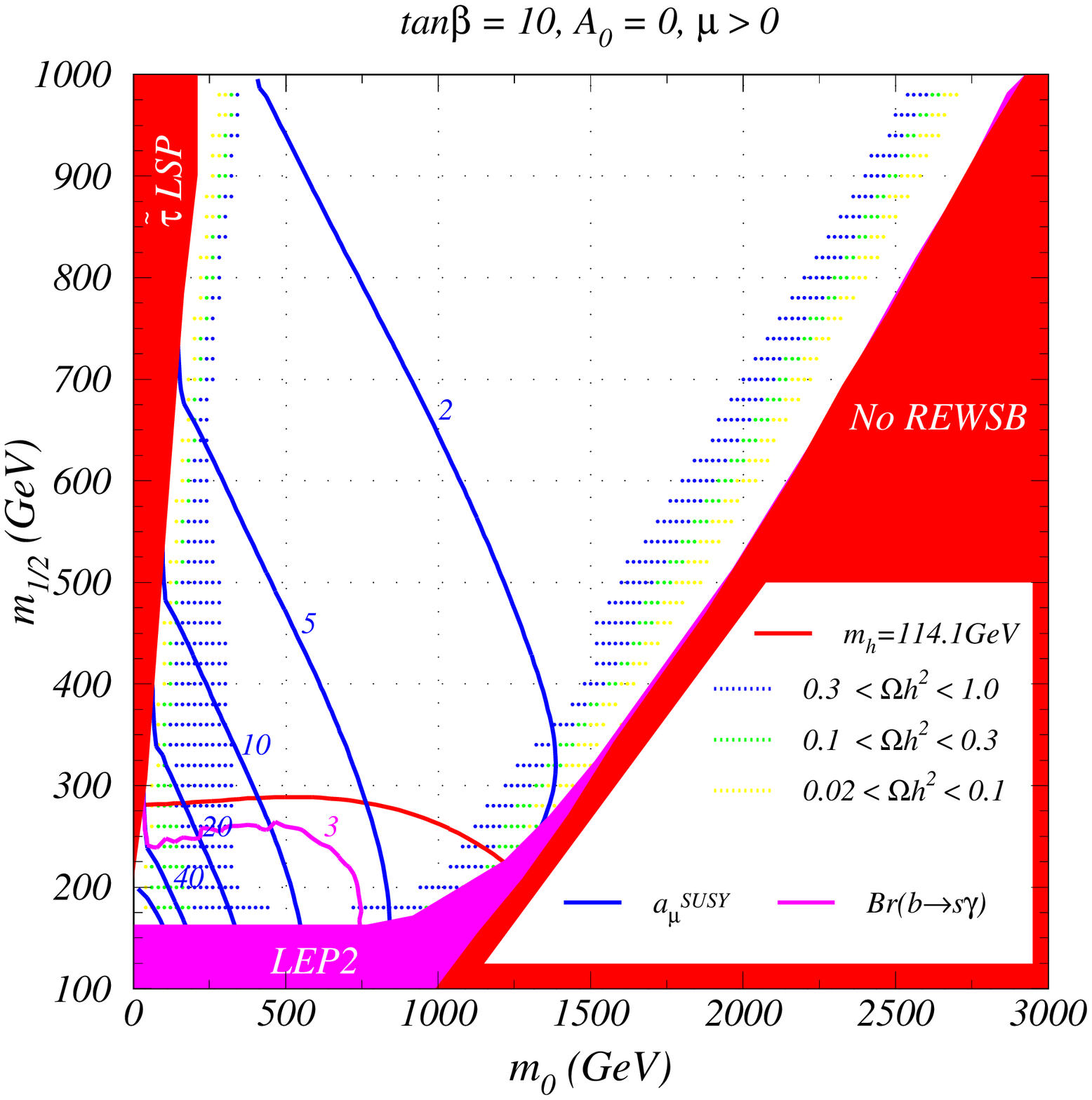,width=15cm} 
\caption{Same as Fig. \ref{fig:sug10m}, but for $\mu >0$.}
\label{fig:sug10p}}

We next turn to the $m_0\ vs.\ m_{1/2}$ plane for $\tan\beta =30$ and
$\mu <0$. The gray region in the bottom left corner of the plot is
excluded because $m_{\ttau_1}^2 < 0$. In this case, the area of the green
region of relic density in the lower-left has expanded considerably
owing to enhanced neutralino annihilation to $b\bar{b}$ and
$\tau\bar{\tau}$ at large $\tan\beta$.  Both lighter values of
$m_{\ttau_1}$ and $m_{\tb_1}$ and also large $\tau$ and $b$ Yukawa
couplings at large $\tan\beta$ enhance these $t$-channel annihilation
rates through virtual staus and sbottoms. Unfortunately, the region
excluded by $BF(b\to s\gamma )$ and by $\delta a_\mu$ (even with the
conservative constraint of Ref.\cite{melnikov}) also expands, and 
most of the cosmologically preferred region is
again ruled out. As before, we are left with the corridors of stau
co-annihilation and the focus point scenario as the only surviving
regions.
 
\FIGURE[t]{\epsfig{file=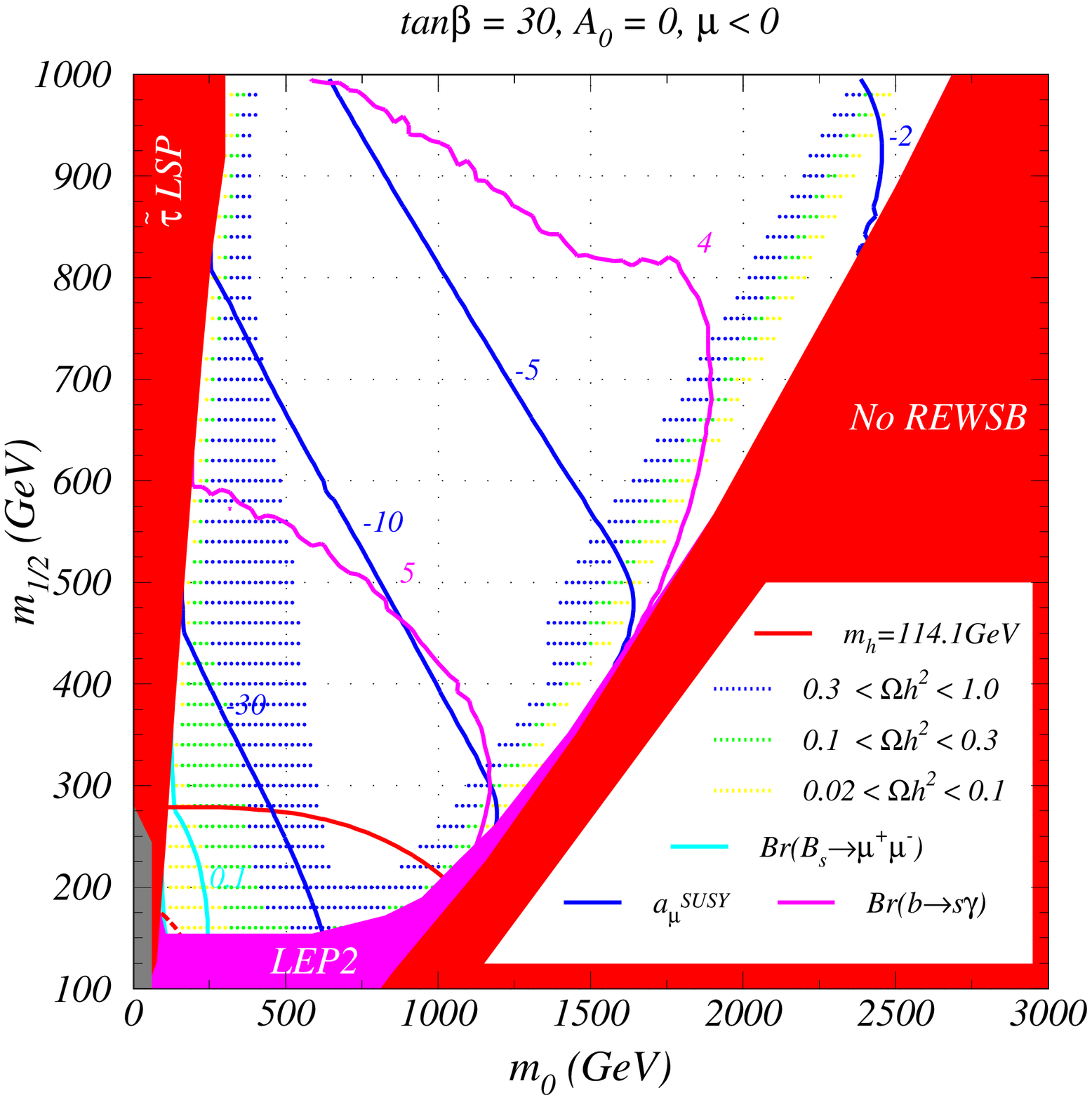,width=15cm} 
\caption{Same as Fig. \ref{fig:sug10m}, but for $\tan\beta =30$ and $\mu
<0$. The light blue contour labeled 0.1 denotes where $B(B_s \to
\mu^+\mu^-)= 0.1 \times 10^{-7}$. In subsequent figures these branching
fractions contours are all labeled in units of $10^{-7}$. }
\label{fig:sug30m}}

The corresponding plot is shown for $\tan\beta =30$ but $\mu >0$
in Figure \ref{fig:sug30p}. In this case, the magenta contours of
$BF(b\to s\gamma )$ correspond to $2$ and $3\times 10^{-4}$. Thus, the lower
left region is excluded since it leads to too {\it low} a value
of $BF(b\to s\gamma )$. The $\delta a_\mu $ contours begin from lower left 
with $60\times 10^{-10}$, then proceed to 40, 20, 10, 5 and $2\times 10^{-10}$.
A fraction of the slepton annihilation region of relic density is
excluded also by too large a value of $\delta a_\mu$. Of course,
a reasonable relic density may also be achieved in the stau co-annihilation 
and focus point regions of parameter space.

\FIGURE[t]{\epsfig{file=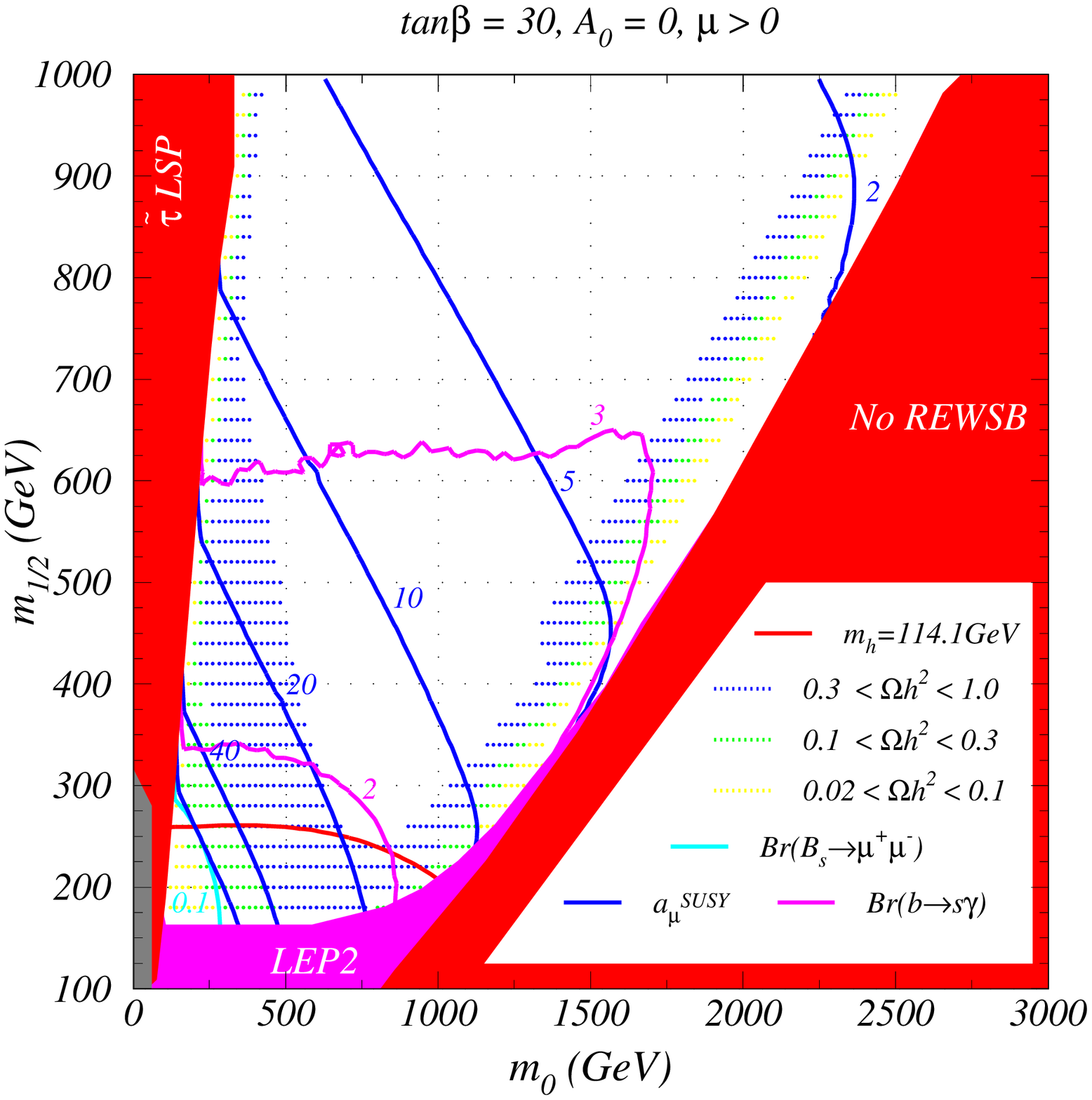,width=15cm} 
\caption{Same as Fig. \ref{fig:sug10m}, but for $\tan\beta =30$ and 
$\mu >0$.}
\label{fig:sug30p}}

Next, we examine the mSUGRA parameter plane for very large values of
$\tan\beta =45$ and $\mu <0$. If we take $\tan\beta$ much bigger than 45
for this sign of $\mu$, the entire parameter space is excluded due to
lack of REWSB. The gray and red regions are as in previous figures. The
blue region is excluded because $m_A^2<0$, denoting again a lack of
appropriate REWSB. The inner and outer red dashed lines are contours of
$m_A=100$ and $m_A= 200$~GeV, respectively. The former is roughly the
lower bound on $m_A$ from LEP experiments. In between these contours,
$h$ is not quite SM-like, and the mass bound from LEP may be somewhat lower
than $m_h=114.1$~GeV shown by the solid red contour, but outside the
200~GeV contour this bound should be valid.

The tiny black region in the figure is excluded by 
constraints\cite{zprop} on the
width of the $Z$ boson (specifically, the decay $Z \to hA+HA$ 
leads to too large a value for $\Gamma_Z$). 
We also see that much of the lower-left region is excluded by too high a
value of $BF(b\to s \gamma )$ and too low a value of $\delta a_\mu$.  In
addition, in this plane, the experimental limit on $B_s\to\mu^+\mu^-$
enters the lower-left, where values exceeding $26\times 10^{-7}$ are
obtained.
It seems that in the upper region which is  favored by the $b \to s\gamma$
constraint, detection of $B_s \to \mu^+\mu^-$ at the Tevatron will be
quite challenging.

In this figure, the relic density regions are qualitatively
different
from the lower $\tan\beta$ plots. A long diagonal strip running from
lower-left to upper-right occurs because in this region, neutralinos
annihilate very efficiently through $s$-channel $A$ and $H$ Higgs
graphs, where the total Higgs widths are very large due to the large $b$
and $\tau$ Yukawa couplings for the high value of $\tan\beta$ in this
plot. Adjacent to this region are yellow and green shaded regions where
neutralino annihilation is still dominated by the $s$-channel Higgs
graphs, but in this case the annihilation is somewhat off-resonance. The
$A$ and $H$ widths are so large, typically $20-60$ GeV, that even if
$|2m_{\tz_1}-m_{A(H)}|$ is relatively large, efficient annihilation can
still take place.  The relic density changes so slowly on the flanks of the
annihilation corridor that little fine tuning of parameters is needed
to achieve a favored value of $\Omega_{\tz_1}h^2$.

\FIGURE[t]{\epsfig{file=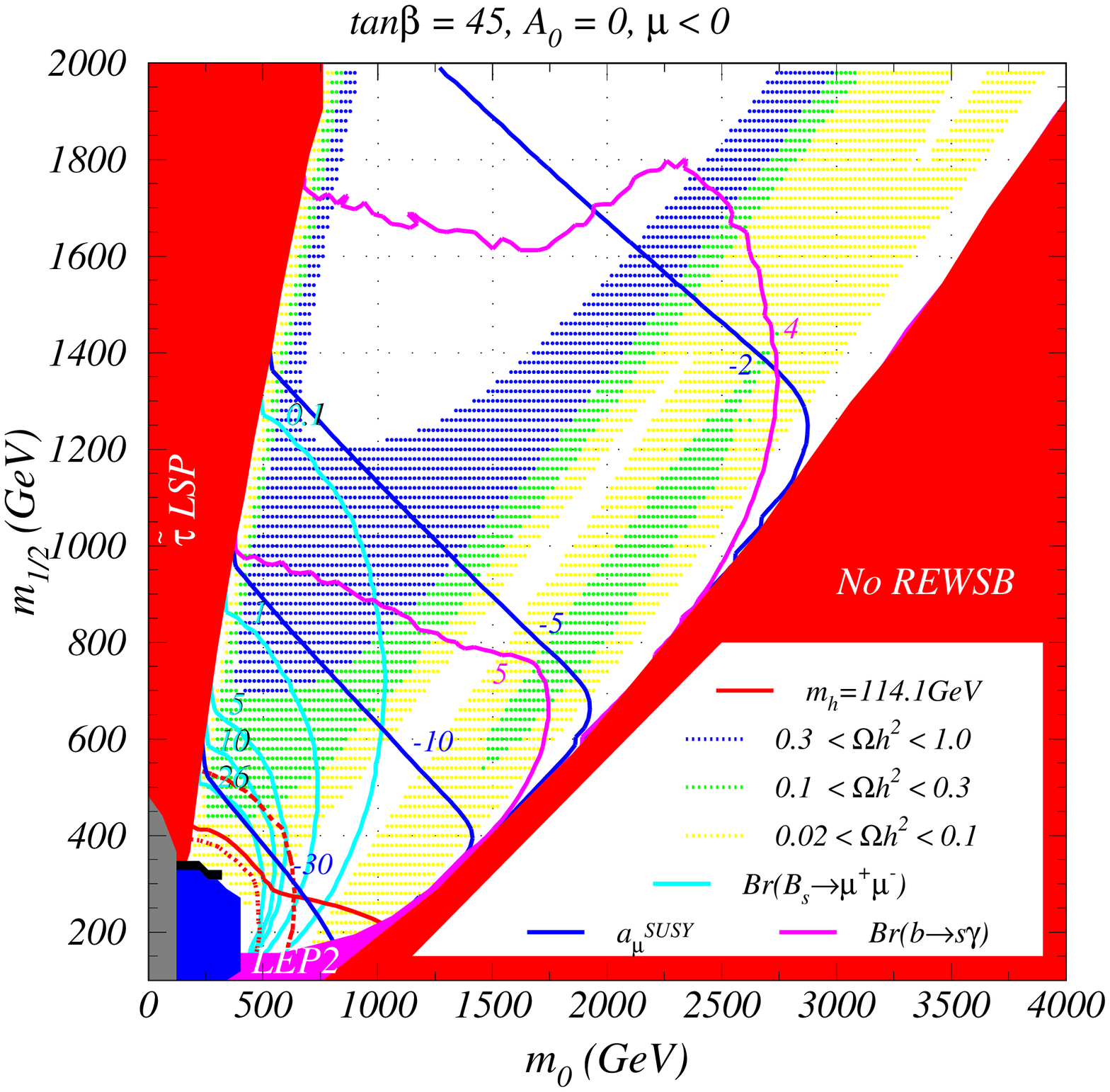,width=15cm} 
\caption{Same as Fig. \ref{fig:sug10m}, but for $\tan\beta =45$ and $\mu <0$.
The inner and outer red dashed lines are contours of $m_A=100$ and 
$m_A= 200$~GeV, respectively.}
\label{fig:sug45m}}

For the case of $\mu >0$, $\tan\beta$
values ranging up to $60$ can be allowed, although the mSUGRA parameter
space becomes very limited for $\tan\beta \agt 55$. 
Hence, we show in Fig. \ref{fig:sug52p} the mSUGRA parameter space
plane for $\tan\beta =52$ and $\mu >0$. 
In this parameter plane, the relic density annihilation corridor 
occurs near the boundary of the excluded $\ttau_1$ LSP region. 
The width of the $A$ and $H$ Higgs scalars is very wide, ranging from 30 GeV
for $m_{1/2}\sim 400$ GeV, to 110-130 GeV for $m_{1/2}\sim 2000$ GeV.
Efficient $s$-channel annihilation through the Higgs poles can occur
throughout much of the upper region of allowed parameter space. But the 
annihilation is not overly efficient due to the extreme breadth of the 
Higgs resonances.
In fact, {\it none} of the entire parameter plane is excluded 
by $\Omega_{\tz_1}h^2>1$!
The change in $\Omega_{\tz_1}h^2$ is so slow over almost the entire
parameter plane that little fine-tuning occurs. 

In much of the region with $m_{1/2}<400$ GeV, the value of
$BF(b\to s\gamma )$ is below
$2\times 10^{-4}$, so that some of the lower green relic density region
where annihilation occurs through $t$-channel stau exchange is excluded.
In contrast, the value of $\delta a_\mu$ is 
in the range of $10-40\times 10^{-10}$, which is in accord with
the E821 measurement. The value of $m_h$ is almost always
above 114.1 GeV, and the $BF(B_s\to\mu^+\mu^- )$ is always
below $10^{-7}$, and could (if at all) be detected with several years of
main injector operation. 

Aside from the somewhat low value of
$BF(b\to s\gamma )$ in the lower left, much of this plane 
represents a very attractive area
of mSUGRA model parameter space. If the model parameters are indeed in
this range, the Tevatron signal for $B_s \to \mu^+\mu^-$ will be
small, and $BF(b\to s\gamma )$ will turn out somewhat below the SM prediction,
while $\delta a_\mu$ will be somewhat above the SM value. 

\FIGURE[t]{\epsfig{file=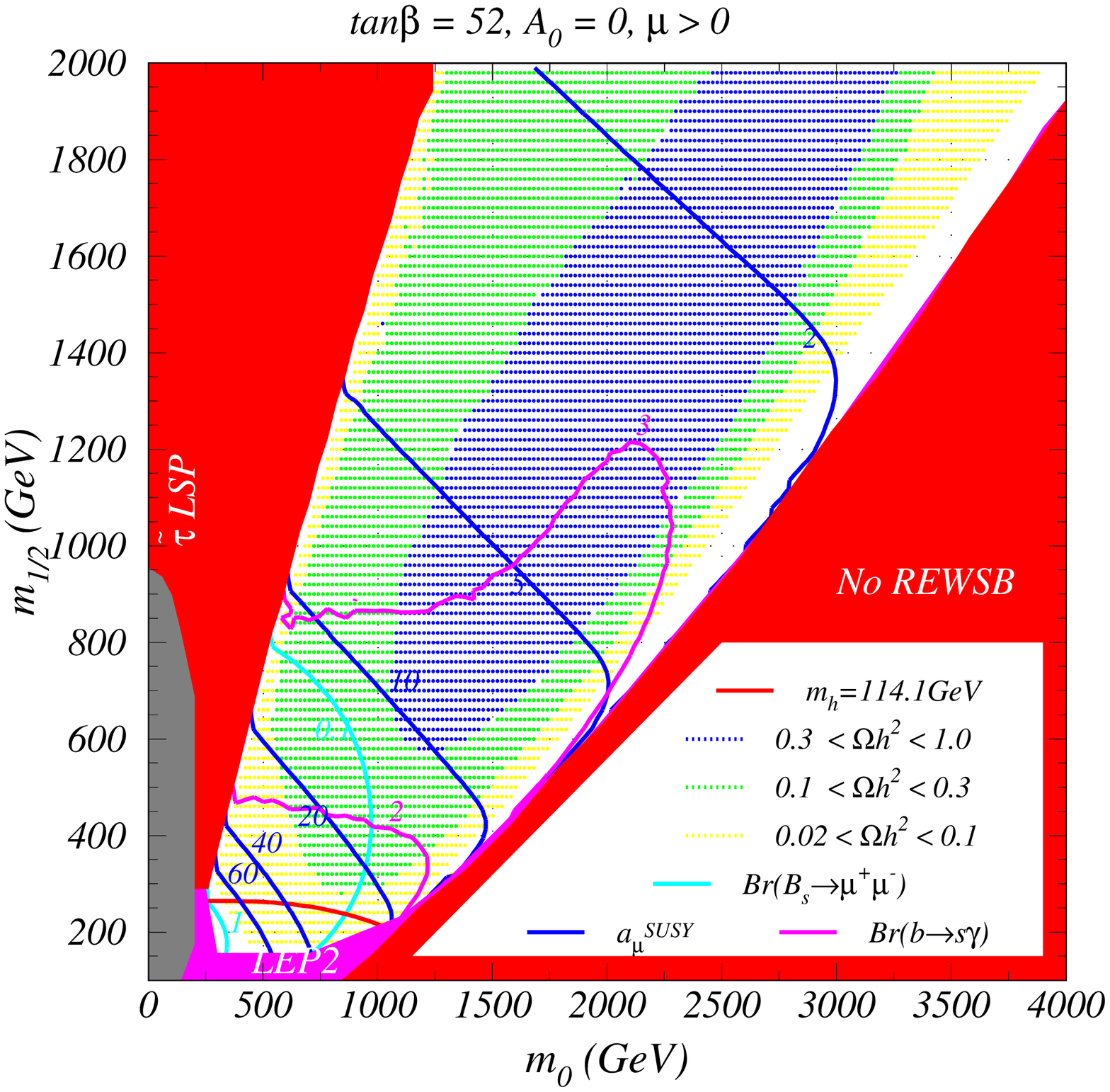,width=15cm} 
\caption{Same as Fig. \ref{fig:sug10m}, but for $\tan\beta =52$ and 
$\mu >0$.}
\label{fig:sug52p}}

\section{Conclusions}
\label{sec:conclude}

In this paper, we have presented updated constraints on the mSUGRA model
from {\it i.}) the LEP2 constraints on sparticle and Higgs boson masses, 
{\it ii.}) the neutralino relic density $\Omega_{\tz_1}h^2$, 
{\it iii.}) the branching fraction $BF(b\to s\gamma )$, 
{\it iv.}) the muon anomalous magnetic moment $a_\mu$ and
{\it v.}) the leptonic decay $B_s\to\mu^+\mu^-$. Putting all five
constraints together, we find favored regions of parameter space which
may be categorized by the mechanism for annihilating relic neutralinos
in the early universe:
\bi
\item {\bf 1.} annihilation through $t$-channel slepton exchange
(low $m_0$ and $m_{1/2}$),
\item {\bf 2.} the stau co-annihilation region 
(very low $m_0$ but large $m_{1/2}$),
\item {\bf 3.} the focus point region (large $m_0$ but low to 
intermediate $m_{1/2}$) and
\item {\bf 4.} the flanks of the neutralino $s$-channel annihilation via
$A$ and $H$ corridor at large $\tan\beta$ when $\Gamma_A$ and $\Gamma_H$
are very large.
\ei

In previous years, there may have been a preference for region {\bf 1.}
as offering the most natural channel for obtaining a reasonable value
of relic density. However, recently much of this region is ruled out by
a combination of LEP2 limits on $m_h$ at low $\tan\beta$, too high 
(for $\mu <0$) or too low (for $\mu >0$ and large $\tan\beta$) a value of
$BF(b\to s\gamma )$, and too low a value of $a_\mu$ (for $\mu <0$ and
intermediate to large $\tan\beta$). In addition, the CDF
$BF(B_s\to \mu^+\mu^-)$ constraint is starting to become important
for $\mu <0$ and large $\tan\beta$. Nevertheless, some of region 1 remains 
viable, especially for $\mu >0$, where we expect a lightest Higgs boson 
just beyond the bounds from LEP2.

The stau co-annihilation region\cite{ellis_co} {\bf 2.} is intriguing 
because one can always take
$m_{1/2}$ large enough for any $\tan\beta$ value to evade constraints
on deviations from SM predictions. However, this region is exceptionally 
narrow in the parameter $m_0$, and slight deviations cause either too high
or too low a value of relic density. This indicates a high degree of 
fine-tuning in the determination of the relic density in this region.

The focus point region\cite{feng} {\bf 3.} also leads to a 
reasonable relic density,
this time because the higgsino component of $\tz_1$ is large enough
that efficient annihilation can occur to $WW$, $ZZ$ and $ZH$ states.
It may also be preferable based on possibly low electroweak fine-tuning, and
because matter scalar masses are high enough to offer some degree of 
suppression of SUSY induced FC and CP violating processes\cite{feng}.
This region also suffers some degree of fine-tuning of the relic density,
since too high or too low a higgsino component of $\tz_1$ can result in
too low or too high a value of relic density. For $\tan\beta\sim 10$ and
$\mu >0$, the focus point region is in accord with all constraints. 
For $\mu <0$, the focus point region usually gives too high a value of
$BF(b\to s\gamma )$; for $\mu >0$, we would expect ultimately
experiment to measure a somewhat lower value of $BF(b\to s\gamma )$ 
than the SM prediction, and a somewhat higher value of $a_\mu$.

Finally, at very large $\tan\beta\sim 45-55$ there can exist wide regions of
parameter space where $\tz_1\tz_1$ annihilation can occur in the early 
universe through very broad $s$-channel $A$ and $H$ resonances,
giving rise to a reasonable value of relic density\cite{dn,bb}: 
region {\bf 4.}. 
Unfortunately, much of 
this region is excluded for $\mu <0$ by too large a value of 
$BF(b\to s\gamma )$. Here also a rather low a value of $a_\mu$ less than the
SM prediction is generated. In this case, very large values of $m_0$ and 
$m_{1/2}$ are needed to escape experimental constraints, perhaps
placing the SUSY spectrum in conflict with naturalness bounds\cite{diego}.

For very large $\tan\beta$ and $\mu >0$, however, broad regions of 
parameter space can be found with a reasonable relic density, and also 
which are in accord with all other low energy constraints. 
In this region, we expect $BF(b\to s\gamma )$ somewhat below the SM 
prediction, and $a_\mu$ somewhat above the SM prediction.
The current $BF(b\to s\gamma )$ measurement suggest 
$m_{1/2}$ values $\agt 500$ GeV, giving rise to sparticles 
typically at the TeV scale or beyond.

To summarize these various regions, we present in Table \ref{table} 
five parameter space points indicating the SUSY spectrum
that might occur in each region. 
We also list the relic density, $BF(b\to s\gamma )$, $a_\mu$ and
$BF(B_s\to\mu^+\mu^- )$ values. 

Point 1 occurs in region {\bf 1.}, and is characterized by light
sparticle masses, especially tau sleptons.  The light Higgs scalar $h$
is slightly beyond the LEP2 bound.  The search channel at the Fermilab
Tevatron would be $p\bar{p}\to\tw_1\tz_2 X$, with $\tz_2\to\tau\ttau_1$
and $\tw_1\to\ttau\nu_\tau$ though detection appears to be
difficult\cite{tevatron}.  The CERN LHC would be awash in
signals\cite{lhc}, and many new SUSY states would be accessible to a
linear collider (LC) with $\sqrt{s}\simeq 500-800$ GeV\cite{nlc}.

Point 2 is in the stau co-annihilation region {\bf 2.}, but at low
enough $m_{1/2}$ that $t$-channel stau exchange is still important in
the $\tz_1\tz_1$ annihilations. It is characterized by a rather small
mass gap between $\ttau_1$ and $\tz_1$. Only the Higgs boson $h$
would be observable at the Tevatron. At the LHC, a variety of 
leptonic signatures would occur in gluino and squark cascade decays, 
including a high rate of $\tau$ production. The LC would have to operate
slightly above $\sqrt{s}\sim 500$ GeV to access even the first SUSY states.

Point 3 in the focus point region {\bf 3.} has TeV scale scalars, but
light $\tw_1$ and $\tz_2$. The LSP is a mixture of gaugino-higgsino.
The $\tw_1$ decays into three bodies dominated by $W^*$ exchange, and the
$\tz_2$ decay is dominated by $Z^*$ exchange. Because this point is in
the region with small $|\mu|$ (but with a not too small
$m_{\tw_1}-m_{\tz_1}$), a low rate of 
trilepton events may be accessible to Fermilab Tevatron experiments.
At the CERN LHC, $\tg\tg$ and $\tw_1\tz_2$ production would be dominant.
A LC would be able to find most of the charginos and neutralinos, since
these are the lowest lying states.

Point 4a lies in region {\bf 4.} on the edge of the Higgs annihilation 
corridor for large $\tan\beta$
and $\mu <0$. To evade constraints, the SUSY spectrum is very heavy:
only the Higgs $h$ would be seen at the Tevatron or a LC, 
and in fact it would even be challenging to discover SUSY at the LHC.
This point is disfavored by naturalness arguments.

Point 4b also in region {\bf 4.} lies in the Higgs annihilation 
corridor for large
$\tan\beta$ and $\mu>0$. A significantly lighter spectrum can be tolerated
than in the 4a case, although SUSY scalars are still in the TeV range.
Only the $h$ would be accessible at the Tevatron. SUSY signals
corresponding to point 4b
should be readily visible at the LHC, although a LC with $\sqrt{s}>1$ TeV
would be required to access even the lowest lying SUSY states.

To summarize, we find the five constraints considered in this paper
to be highly restrictive. Together, they rule out large regions of parameter
space of the mSUGRA model, including much of the region where
$t$-channel slepton annihilation of neutralinos occurs in the early universe.
The surviving regions {\bf 1.}-{\bf 4.} have distinct characteristics of their
SUSY spectrum, and should lead to distinct SUSY signatures at colliders.

\begin{table}
\begin{center}
\caption{Representative weak scale sparticle masses 
(in GeV units) and parameters for five selected mSUGRA models.
 We use $A_0=0$ and $m_t=175$ GeV. The value of $\mu$ is also
shown since it is sometimes regarded as a measure of fine tuning. }
\bigskip
\begin{tabular}{lrrrrr}
\hline
parameter & \multicolumn{5}{c}{value}  \\
\hline
 $point$           &   1   &    2   &    3   &    4a  &    4b  \\
 $m_0$             &   100 &    165 &   1200 &   2750 &    800 \\
 $m_{1/2}$         &   300 &    550 &    250 &   1800 &    800 \\
 $\tan(\beta)$     &    10 &     10 &     10 &     45 &     52 \\
 $sgn(\mu)$        &     1 &      1 &      1 &   $-$1 &      1 \\
\hline
 $m_{\tg}$         & 701.4 & 1225.6 &  658.0 & 3810.8 & 1757.6 \\
 $m_{\tu_L}$       & 630.7 & 1099.1 & 1271.1 & 4185.0 & 1715.3 \\
 $m_{\tu_R}$       & 611.1 & 1060.0 & 1269.1 & 4080.7 & 1662.4 \\
 $m_{\td_L}$       & 635.6 & 1102.0 & 1273.6 & 4185.8 & 1717.1 \\
 $m_{\td_R}$       & 610.0 & 1055.7 & 1269.6 & 4067.4 & 1656.5 \\
 $m_{\tb_1}$       & 584.9 & 1020.8 & 1072.4 & 3501.7 & 1484.4 \\
 $m_{\tb_2}$       & 610.7 & 1053.4 & 1260.8 & 3537.6 & 1539.4 \\
 $m_{\tst_1}$      & 471.7 &  858.5 &  825.2 & 3213.2 & 1328.2 \\
 $m_{\tst_2}$      & 648.1 & 1064.0 & 1084.3 & 3529.6 & 1533.6 \\
 $m_{\tnu_{e}}$    & 216.4 &  396.7 & 1203.1 & 2972.3 &  952.4 \\
 $m_{\te_L}$       & 230.4 &  404.5 & 1205.7 & 2973.4 &  955.8 \\
 $m_{\te_R}$       & 155.5 &  264.8 & 1201.7 & 2822.1 &  851.7 \\
 $m_{\tnu_{\tau}}$ & 215.6 &  395.4 & 1198.0 & 2750.8 &  834.0 \\
 $m_{\ttau_1}$     & 147.5 &  257.6 & 1191.0 & 2320.8 &  524.2 \\
 $m_{\ttau_2}$     & 233.4 &  405.2 & 1200.8 & 2752.6 &  847.9 \\
 $m_{\tz_1}$       & 117.5 &  225.1 &   88.6 &  785.0 &  336.2 \\
 $m_{\tz_2}$       & 215.1 &  416.9 &  144.1 & 1235.1 &  620.2 \\
 $m_{\tz_3}$       & 398.5 &  668.1 &  198.2 & 1249.2 &  848.7 \\
 $m_{\tz_4}$       & 417.8 &  682.8 &  260.9 & 1461.7 &  862.3 \\
 $m_{\tw_1}$       & 214.7 &  416.9 &  136.5 & 1234.5 &  620.3 \\
 $m_{\tw_2}$       & 418.0 &  682.6 &  260.3 & 1461.7 &  862.5 \\
 $m_h      $       & 114.7 &  119.0 &  114.4 &  123.9 &  121.5 \\
 $m_H      $       & 443.9 &  766.5 & 1204.9 & 1495.2 &  810.1 \\
 $m_A      $       & 443.3 &  765.7 & 1203.9 & 1494.2 &  809.5 \\
 $m_{H^+}  $       & 450.7 &  770.4 & 1207.4 & 1498.4 &  816.5 \\
 $\mu$             & 392.0 &  664.9 & 188.7  & -1247.2&  845.8 \\
 $\Omega h^2$      & 0.232 &  0.218 &  0.262 &  0.210 &  0.181 \\

 $BF(b\to s\gamma)\times 10^4$       
                   &  3.12 &   3.46 &   3.20 &   3.92 &   2.85 \\
 $a_\mu^{SUSY}\times 10^{10}$         
                   &  22.6 &   7.13 &   2.65 &$-$1.48 &   10.2 \\
 $BF(B_s\to\mu^+\mu^-)\times 10^7$   
                   & 0.0399& 0.0389 & 0.0384 & 0.0306 & 0.0870 \\

\hline
\label{table}
\end{tabular}  
\end{center}   
\end{table}    

\section*{Acknowledgments}
 
This research was supported in part by the U.S. Department of Energy
under contracts number DE-FG02-97ER41022 and DE-FG03-94ER40833, and by 
Funda\c{c}\~ao de Amparo \`a Pesquisa do Estado de S\~ao Paulo
(FAPESP).
	
%


\begin{thebibliography}{99}

\bibitem{reviews} For recent reviews, see {\it e.g.} S.~Martin,
in {\it Perspectives on Supersymmetry}, edited by G.~Kane (World Scientific),
\hepph{9709356}; M.~Drees, \hepph{9611409};
J.~Bagger, \hepph{9604232};
X.~Tata, {\it Proc.~IX J.~Swieca Summer School,}
J.~Barata, A.~Malbousson and S.~Novaes, Eds., \hepph{9706307};
S.~Dawson, {Proc. TASI 97}, J.~Bagger, Ed., \hepph{9712464}.

\bibitem{sugra} A.~Chamseddine, R.~Arnowitt and P.~Nath, 
\prl{49}{1982}{970};
R.~Barbieri, S.~Ferrara and C.~Savoy, 
\plb{119}{1982}{343};
L.~J.~Hall, J.~Lykken and S.~Weinberg, \prd{27}{1983}{2359};
for a review, see H.~P.~Nilles, \prep{110}{1984}{1}.

\bibitem{gmsb}
M.~Dine and A.~E.~Nelson, \prd{48}{1993}{1277}; \\
M.~Dine, A.~E.~Nelson and Y.~Shirman, \prd{51}{1995}{1362}; \\
M.~Dine, A.~E.~Nelson, Y.~Nir and Y.~Shirman, \prd{53}{1996}{2658};\\
for a review, see G.~F.~Giudice and R.~Rattazzi, \prep{322}{1999}{419}.

\bibitem{amsb}
L.~Randall and R.~Sundrum, \npb{557}{1999}{79};
G.~F.~Giudice, M.~A.~Luty, H.~Murayama and R.~Rattazzi, \jhep{12}{1998}{027}.

\bibitem{inomsb}
D.~E.~Kaplan, G.~D.~Kribs and M.~Schmaltz, \prd{62}{2000}{035010}; \\
Z.~Chacko, M.~A.~Luty, A.~E.~Nelson and E.~Ponton, \jhep{0001}{2000}{003}.

\bibitem{sw} S.~K.~Soni and H.~A.~Weldon, \plb{126}{1983}{215};
V.~Kaplunovsky and J.~Louis, \plb{306}{1993}{269}.

\bibitem{munoz} K.~Choi, J.~S.~Lee and C.~Munoz, \prl{80}{1998}{3686}.

\bibitem{isajet} H.~Baer, F.~Paige, S.~Protopopescu and X.~Tata,
\hepph{0001086}.

\bibitem{FHF} S.~Heinemeyer, W.~Hollik and G.~Weiglein, 
\hepph{0002213}.

\bibitem{bagger} D.~Pierce, J.~Bagger, K.~Matchev and R.~Zhang,
\npb{491}{1997}{3}.

\bibitem{Gomez:2002tj} M.~E.~Gomez, G.~Lazarides and C.~Pallis, 
\hepph{0203131}.

\bibitem{ellis} J.~Ellis, K.~Olive and Y.~Santoso, \hepph{0202110}.

\bibitem{leszek} L.~Roszkowski, R.~Ruiz de Austri and T.~Nihei,
\jhep{0108}{2001}{024}.

\bibitem{drees} A.~Djouadi, M.~Drees and J.~L.~Kneur,
\jhep{0108}{2001}{055}.

\bibitem{deboer} W.~de Boer, M.~Huber, C.~Sander and D.~I.~Kazakov,
\plb{515}{2001}{283}.

\bibitem{kraml} S.~Kraml, talk given at SUSY 2002, 
\href{http://www.desy.de/~susy02/pa.1b/kraml_susy.ps.gz}
{http://www.desy.de/~susy02/pa.1b/kraml\_susy.ps.gz}.

\bibitem{bbb} H.~Baer, C.~Bal\'azs and A.~Belyaev, 
\jhep{0203}{2002}{042}.

\bibitem{bb}  H.~Baer and M.~Brhlik, \prd{53}{1996}{597} and
\prd{57}{1998}{576}; V.~Barger and C.~Kao, \prd{57}{1998}{3131}
and \plb{518}{2001}{117};
DarkSUSY, by P.~Gondolo and J.~Edsj\"o, \astroph{0012234};
H.~Baer, M.~Brhlik, M.~A.~Diaz, J.~Ferrandis, 
P.~Mercadante, P.~Quintana and X.~Tata,
\prd{63}{2001}{015007}.

\bibitem{belanger} G.~Belanger, F.~Boudjema, A.~Pukhov and A.~Semenov,
\hepph{0112278}.
%
\bibitem{dn} M.~Drees and M.~Nojiri, \prd{47}{1993}{376}.
%
\bibitem{an} P. Nath and R. Arnowitt, \prl{70}{1993}{3696};
R. Arnowitt and P. Nath, \plb{437}{1998}{344}.
%

\bibitem{ellis_ltb} J.~R.~Ellis, T.~Falk, G.~Ganis, K.~A.~Olive and 
M.~Srednicki, \plb{510}{2001}{236}.
%
\bibitem{nano} A.B.~Lahanas and V.C. Spanos,
\epjc{23}{2002}{185}; see also, A.B.~Lahanas,  
D.V.~Nanopoulos and V.C.~Spanos, \prd{62}{2000}{023515}.

\bibitem{cdf} F.~Abe {\em et al.}, (CDF Collaboration), \prd{57}{1998}{3811}.

\bibitem{bmm} K.~Babu and C.~Kolda, \prl{84}{2000}{228};
A.~Dedes, H.~Dreiner and U.~Nierste, \prl{87}{2001}{251804};
R.~Arnowitt, B.~Dutta, T.~Kamon and M.~Tanaka, \hepph{0203069}.

\bibitem{lep2_w1} Joint HEP2 Supersymmetry Working Group, {\it Combined 
Chargino Results, up to 208 GeV},\break
\verb|http://alephwww.cern.ch/lepsusy/www/inos_moriond01/charginos.pub.html.|

\bibitem{lep2_sel} Joint HEP2 Supersymmetry Working Group, {\it Combined 
LEP Selectron/Smuon/Stau Results, 183-208 GeV},\break
\verb|http://alephwww.cern.ch/~ganis/SUSYWG/SLEP/sleptons_2k01.html.|

\bibitem{lep2_h} LEP Higgs Working Group Collaboration, 
\hepex{0107030}.

\bibitem{cdm}  See {\it e.g.} W.~L.~Freedman, 
\pr{333}{2000}{13}.

\bibitem{comphep} CompHEP~v.33.23, by A.~Pukhov {\it et al.}, 
\hepph{9908288}.

\bibitem{graciela} P.~Gondolo and G.~Gelmini, \npb{360}{1991}{145};
J.~Edsj\"o and P.~Gondolo, \prd{56}{1997}{1879}.

\bibitem{belle} K.~Abe {\it et al.} (Belle Collaboration), 
\plb{511}{2001}{151}.

\bibitem{cleo} D.~Cronin-Hennessy {\it et al.} (Cleo Collaboration),
\prl{87}{2001}{251808}.

\bibitem{aleph} R.~Barate {\it et al.} (Aleph Collaboration),
\plb{429}{1998}{169}.

\bibitem{Gambino:2001ew} P.~Gambino and M.~Misiak, 
\npb{611}{2001}{338}.

\bibitem{bsg} H.~Baer and M.~Brhlik, \prd{55}{1997}{3201};
H.~Baer, M.~Brhlik, D.~Casta\~no and X.~Tata, \prd{58}{1998}{015007}.

\bibitem{anlauf} H.~Anlauf, \npb{430}{1994}{245}.

\bibitem{degrassi} G.~Degrassi, P.~Gambino and G.~Giudice,
\jhep{0012}{2000}{009}.

\bibitem{carena} M.~Carena, D.~Garcia, U.~Nierste and C.~Wagner,
\plb{499}{2001}{141}.

\bibitem{e821} H.~N.~Brown {\it et al.}, (E821 Collaboration),
\prl{86}{2001}{2227}.

\bibitem{marciano} A.~Czarnecki and W.~Marciano,
\prd{64}{2001}{013014}. This uses the analysis of M.~Davier and
A.~H\"ocker, \plb{435}{1998}{427} which in turn uses $\tau$ decay
data to reduce the error on the hadronic vacuum polarization. See also
S.~Narison, \plb{513}{2001}{53}. 

\bibitem{lbl} M.~Knecht and A.~Nyffeler, \prd{65}{2002}{073034};
M.~Knecht, A.~Nyffeler, M.~Perrottet and E.~De Rafael, 
\prl{88}{2002}{071802};
M.~Hayakawa and T.~Kinoshita, \hepph{0112102};
I.~Blokland, A.~Czarnecki and K.~Melnikov,
\prl{88}{2002}{071803}. 

\bibitem{melnikov} K.~Melnikov, \ijmpa{16}{2001}{4591}. His updated
analysis of the SM value of $\delta a_{\mu}$ was presented at the High
Energy Physics Seminar, University of Hawaii, March 2002.

\bibitem{jag} F.~Jegerlehner, \hepph{0104304}.

\bibitem{bbft} H.~Baer, C.~Bal\'azs, J.~Ferrandis and X.~Tata,
\prd{64}{2001}{035004}.

\bibitem{mtw} J.~K.~Mizukoshi, X.~Tata and Y.~Wang (in preparation). 

\bibitem{ellis_co} J.~Ellis, T.~Falk and K.~Olive, \plb{444}{1998}{367};
J.~Ellis, T.~Falk, K.~Olive and M.~Srednicki, 
\app{13}{2000}{181}.

\bibitem{ellis_ft} J.~Ellis and K.~Olive, \plb{514}{2002}{114}.

\bibitem{feng_relic} J.~Feng, K.~Matchev and F.~Wilczek,
\plb{482}{2000}{388} and \prd{63}{2001}{045024}.

\bibitem{feng} J.~Feng, K.~Matchev and T.~Moroi, 
\prd{61}{2000}{075005}.

\bibitem{run2} M.~Carena {\em et al.}, \hepph{0010338}.

\bibitem{zprop} H.~Baer, M.~Drees and X.~Tata, \prd{41}{1990}{3414};
J.~Ellis, G.~Ridolfi and F.~Zwirner, \plb{237}{1990}{423}.

\bibitem{diego} G.~W.~Anderson and D.~J.~Casta\~no, \plb{347}{1995}{300}
and \prd{52}{1995}{1693}.
%
\bibitem{tevatron} H.~Baer, C.~H.~Chen, M.~Drees, F.~Paige and X.~Tata,
\prd{58}{1998}{075008}; V.~Barger, C.~Kao and T.~Li, \plb{433}{1998}{328};
V.~Barger and C.~Kao, \prd{60}{1999}{115015};
K.~Matchev and D.~Pierce, \prd{60}{1999}{075004} and
\plb{467}{1999}{225}; J.~Lykken and K.~Matchev, \prd{61}{2000}{015001};
H.~Baer, M.~Drees, F.~Paige, P.~Quintana and X.~Tata, \prd{61}{2000}{095007};
for a review, see S.~Abel {\it et al.}  (SUGRA Working Group Collaboration),
\hepph{0003154}.
%
\bibitem{lhc} H.~Baer, C.~H.~Chen, F.~Paige and X.~Tata,
\prd{52}{1995}{2746} and \prd{53}{1996}{6241};
H.~Baer, C.~H.~Chen, M.~Drees, F.~Paige and X.~Tata, \prd{59}{1998}{055014};
S.~Abdullin {\it et al.}  (CMS Collaboration), \hepph{9806366}.
%
\bibitem{nlc} H.~Baer, R.~Munroe and X.~Tata, \prd{54}{1996}{6735};
for a review, see T.~Abe {\it et al.}  (American Linear Collider 
Working Group Collaboration), \hepex{0106056};
J.~A.~Aguilar-Saavedra {\it et al.}  (ECFA/DESY LC Physics Working Group
Collaboration), \hepph{0106315}.

%


\end{thebibliography}
\end{document}